\documentclass[10pt]{article}
\begin{document}
\begin{center}
{\large\bf Large Scale Cosmological Inhomogeneities, Inflation And
Acceleration Without Dark Energy } \vskip 0.3 true in {\large J.
W. Moffat} \vskip 0.3 true in {\it The Perimeter Institute for
Theoretical Physics, Waterloo, Ontario, N2J 2W9, Canada} \vskip
0.3 true in and \vskip 0.3 true in {\it Department of Physics,
University of Waterloo, Waterloo, Ontario N2Y 2L5, Canada}
\end{center}
\begin{abstract}%
We describe the universe as a local, inhomogeneous spherical
bubble embedded in a flat matter dominated FLRW universe.
Generalized exact Friedmann equations describe the expansion of
the universe and an early universe inflationary de Sitter solution
is obtained. A non-perturbative expression for the deceleration
parameter $q$ is derived that can possibly describe the
acceleration of the universe without dark energy, due to the
effects associated with very long wave length super-horizon
inflationary perturbations. The suggestion by Kolbe et
al.~\cite{Kolbe} that long wave length super-horizon inflationary
modes can affect a local observable through inhomogeneities is
considered in the light of our exact inhomogeneous model.
\end{abstract}
\vskip 0.2 true in
e-mail: jmoffat@perimeterinstitute.ca


\section{Introduction}

In a recent article, we investigated a cosmology in which a
spherically symmetric perturbation enhancement is embedded in an
asymptotic FLRW universe~\cite{Moffat}. The perturbation
enhancement is described by an exact inhomogeneous solution of
Einstein's field equations. We found that the large-scale
inhomogeneities can lead to a reinterpretation of the luminosity
distance $d_L$ of a cosmological source in terms of its red shift
$z$, owing to the observer dependence of these quantities. The
time evolution and the expansion rate of the inhomogeneous
universe can lead to intrinsic effects such as cosmic variance at
large angles and long-wavelength perturbations not described by a
FLRW homogeneous and isotropic universe. Therefore, the
interpretation of the data using a FLRW model that the
accelerating expansion of the universe is caused by dark energy
may be misleading. This is important, for it is difficult to
explain theoretically the postulated dark energy that causes the
acceleration of the universe. The model also leads to an axis
pointing towards the center of the spherically symmetric large
scale perturbation enhancement with dipole, quadrupole and
octopole moments aligned with the axis. It was shown that the
luminosity distances and red shifts observed by different
observers located at spatially different points of causally
disconnected parts of the universe can have varying values. A
spatial average of all these observations leads to an intrinsic
cosmic variance in e.g. the deceleration parameter $q$. The
distribution of CMB temperature fluctuations can be unevenly
distributed in the northern and southern hemispheres.

The popular explanation for the observed large-scale homogeneity
of the universe is that the universe underwent an initial
inflationary period with more than 60 e-folds~\cite{Guth}. The
inflationary cosmic expansion can stretch an initially small,
smooth spatial region to a size larger than the horizon size today
and explain the present day large-scale homogeneity. The question
arises as to whether the initial {\it local} patch can be
sufficiently homogenized to allow inflation to
begin~\cite{Trodden}. In the following, we shall consider the
universe as an expanding bubble with an inhomogeneous metric and
generalized Friedmann equations, including inhomogeneous density
and pressure and a cosmological constant. Our main assumptions are
spherical symmetry and an inhomogeneous barytropic fluid that
satisfies an equation of state. For the case of a spatially flat
inhomogeneous early universe, we obtain a de Sitter inflationary
solution.

The acceleration of the expansion of the universe deduced from
Type Ia supernovae observations and the CMB WMAP
data~\cite{Perlmutter,Riess,Spergel} has been interpreted as due
to the cosmological constant (vacuum energy), modifications of
Einstein's gravitational field equations at large
distances~\cite{Turner}, and quintessence fields~\cite{Peebles}.
The quintessence explanations postulate a new form of matter with
negative pressure called dark energy. Recently, it has been
suggested that the acceleration is caused by very long wavelength,
super-horizon perturbations generated by a period of inflation in
the early universe~\cite{Kolbe,Barausse}. The backreaction of
perturbations on an FLRW background universe has been the subject
of investigation by several authors~\cite{Brandenberger}. The
predictions based on perturbation theory are limited by the
condition $\Phi\ll 1$, where $\Phi$ is the gravitational
potential.

The inflationary perturbation modes whose wavelengths presently
are smaller than $\lambda\leq 10$ Mpc have entered the non-linear
regime and have generated galaxies and clusters of galaxies, while
longer wavelength modes at super-horizon scales $\geq c/H$ are
entering the linear regime today. The effects of the sub-horizon
modes are small due to fact that $\delta\rho/\rho\sim 10^{-5}$ at
the surface of last scattering. Therefore, these sub-horizon modes
produce negligible corrections at second order $\sim 10^{-8}$.
However, the super-horizon modes could potentially create a
correction to the deceleration parameter $q$, large enough to
remove the need for dark energy. It has been argued
recently~\cite{Chung,Flanagan,Seljak,Wiltshire} that second order
perturbation effects of the form $\Phi\nabla^2\Phi$ are described
by a renormalization of the local spatial curvature and cannot
(for a positive energy density) produce a negative deceleration
parameter.

One problem with the perturbation calculations is that they ignore
all higher gradient terms $\nabla^n\Phi$, and any {\it
non-perturbative effects} that can have a significant influence on
the inhomogeneity contributions due to very long wave length modes
at super-horizon. These effects will occur for inflationary models
in which the number of e-folds of inflation is much larger than
the 60 e-folds required to create physically satisfactory
fluctuations. In the following, we shall use the exact
inhomogeneous model of ref.~\cite{Moffat} to derive a formula for
the deceleration parameter $q$ that is non-perturbative and whose
variance can lead to a negative value for $q$ without dark energy
and a cosmological constant $\Lambda$.

\section{Inhomogeneous Friedmann Equations}

Our action takes the form
\begin{equation}
S=S_G+S_M,
\end{equation}
where
\begin{equation}
S_G=\frac{1}{16\pi G}\int d^4x\sqrt{-g}(R-2\Lambda).
\end{equation}
The matter action is given by
\begin{equation}
S_M=\int
d^4x\sqrt{-g}[\frac{1}{2}g^{\mu\nu}\partial_\mu\phi\partial_\nu\phi-V(\phi)],
\end{equation}
where $\phi$ is a scalar matter field and $V(\phi)$ is a
potential.

For the sake of notational clarity, we write the FLRW line element
\begin{equation}
ds^2=dt^2-a^2(t)\biggl(\frac{dr^2}{1-kr^2}+r^2d\Omega^2\biggr),
\end{equation}
where $d\Omega^2=d\theta^2+\sin\theta^2d\phi^2$. The general,
spherically symmetric inhomogeneous line element is given
by~\cite{Lemaitre,Tolman,Bondi,Bonnor,Moffat2,Moffat3,Krasinski,Moffat}:
\begin{equation}
\label{inhomometric} ds^2=dt^2-X^2(r,t)dr^2-R^2(r,t)d\Omega^2.
\end{equation}
The energy-momentum tensor ${T^\mu}_\nu$ takes the barytropic form
\begin{equation}
\label{energymomentum} {T^\mu}_\nu=(\rho+p)u^\mu u_\nu
-p{\delta^\mu}_\nu,
\end{equation}
where $u^\mu=dx^\mu/ds$ and, in general, the density
$\rho=\rho(r,t)$ and the pressure $p=p(r,t)$ depend on both $r$
and $t$. We have for comoving coordinates $u^0=1, u^i=0,\,
(i=1,2,3)$ and $g^{\mu\nu}u_\mu u_\nu=1$.

The Einstein gravitational equations are
\begin{equation}
\label{Einstein} G_{\mu\nu}\equiv
R_{\mu\nu}-\frac{1}{2}g_{\mu\nu}{\cal R}+\Lambda g_{\mu\nu}=-8\pi
GT_{\mu\nu},
\end{equation}
where ${\cal R}=g^{\mu\nu}R_{\mu\nu}$ and $\Lambda$ is the
cosmological constant. Solving the $G_{01}=0$ equation for the
metric (\ref{inhomometric}), we find that
\begin{equation}
X(r,t)=\frac{R'(r,t)}{f(r)},
\end{equation}
where $R'=\partial R/\partial r$ and $f(r)$ is an arbitrary
function of $r$.

We obtain the two generalized Friedmann equations~\cite{Moffat}:
\begin{equation}
\label{inhomoFriedmann} \frac{{\dot R}^2}{R^2}+2\frac{{\dot
R}'}{R'}\frac{{\dot R}}{R}+\frac{1}{R^2}(1-f^2)
-2\frac{ff'}{R'R}=8\pi G\rho+\Lambda,
\end{equation}
\begin{equation}
\label{inhomoFriedmann2} \frac{\ddot
R}{R}+\frac{1}{3}\frac{\dot{R}^2}{R^2}
+\frac{1}{3}\frac{1}{R^2}(1-f^2) -\frac{1}{3}\frac{{\dot
R}'}{R'}\frac{{\dot R}}{R}+\frac{1}{3} \frac{R''ff'}{R'^2R}
=-\frac{4\pi G}{3}(\rho+3p)+\frac{1}{3}\Lambda,
\end{equation}
where $\dot R=\partial R/\partial t$.

\section{De Sitter Inflationary Solution}

Let us now consider the very early universe and retain the
pressure $p$ and the cosmological constant $\Lambda$. We shall
picture the early universe as an expanding spherically symmetric
bubble with its origin at the big bang. We choose $f(r)=1$ for all
values of $r$, and obtain the generalized Friedmann equations
\begin{equation}
\label{inhomoF} \frac{{\dot R}^2}{R^2}+2\frac{\dot R}{R}\frac{\dot
R'}{R'}=8\pi G\rho+\Lambda,
\end{equation},
\begin{equation}
\label{inhomoF2} \frac{\ddot {R}}{R}+\frac{1}{3}\frac{{\dot
R}^2}{R^2}-\frac{1}{3}\frac{{\dot R}'}{R'}
\frac{\dot{R}}{R}=-\frac{4\pi G}{3}(\rho+3p)+\frac{1}{3}\Lambda.
\end{equation}
By using the notation $H_\perp=\dot{R}/R$ and $H_r=\dot{R}'/R'$,
we can write (\ref{inhomoF}) and (\ref{inhomoF2}) as
\begin{equation}
H^2_\perp+2H_\perp H_r=8\pi G\rho+\Lambda,
\end{equation}
\begin{equation}
\frac{\ddot{R}}{R}+\frac{1}{3}H_\perp^2-\frac{1}{3}H_r
H_\perp=-\frac{4\pi G}{3}(\rho+3p)+\frac{1}{3}\Lambda.
\end{equation}
For $H_\perp(r,t)=H_r(r,t)=H(t)=\dot{a}(t)/a(t)$ and
$R(r,t)=a(t)$, we obtain the Friedmann equations of FLRW for a
spatially flat universe:
\begin{equation}
H^2=\frac{8\pi G\rho}{3}+\frac{1}{3}\Lambda.
\end{equation}
\begin{equation}
\frac{\ddot{a}}{a}=-\frac{4\pi G}{3}(\rho+3p)+\frac{1}{3}\Lambda,
\end{equation}
where $\rho=\rho(t)$ and $p=p(t)$.

From the Bianchi identities $\nabla_\nu G^{\mu\nu}=0$, we obtain
for the inhomogeneous model the conservation law
\begin{equation}
\nabla_\nu T^{\mu\nu}=0.
\end{equation}
This becomes
\begin{equation}
\label{pequation}
\partial_\nu
p+\frac{1}{\sqrt{-g}}\partial_\nu[\sqrt{-g}(\rho+p)u^\mu
u^\nu]+{\Gamma^\mu}_{\nu\lambda}u^\nu u^\lambda=0.
\end{equation}
For ${\Gamma^\mu}_{00}=0$ and
$\sqrt{-g}=X(r,t)R^2(r,t)\sin\theta$, we obtain
\begin{equation}
\frac{dp}{dt}=\frac{1}{XR^2}\frac{d}{dt}[XR^2(\rho+p)].
\end{equation}

From (\ref{energymomentum}) we get
\begin{equation}
\label{rhoequation}
\rho_\phi=\frac{1}{2}{\dot\phi}^2+\frac{1}{2}(\vec\nabla\phi)^2+V(\phi),
\end{equation}
\begin{equation}
p_\phi=\frac{1}{2}{\dot\phi}^2-\frac{1}{6}(\vec\nabla\phi)^2-V(\phi).
\end{equation}
Here, $\phi=\phi(r,t)$ depends on both $r$ and $t$. From
(\ref{pequation}) we obtain
\begin{equation}
{\dot\rho_\phi}+\frac{1}{XR^2}\frac{d}{dt}(XR^2)(\rho_\phi+p_\phi)=0.
\end{equation}
Differentiating (\ref{rhoequation}) with respect to $t$ we have
\begin{equation}
\label{waveequation}
{\ddot\phi}+\frac{1}{2}\frac{\partial}{\partial\phi}(\vec\nabla\phi)^2
+\frac{\partial}{\partial\phi}V(\phi)+\frac{d/dt(XR^2)}{XR^2}[\dot\phi
+\frac{1}{3}(\vec\nabla\phi)^2/{\dot\phi}]=0.
\end{equation}

Let us assume an equation of state
\begin{equation}
p(r,t)=w\rho(r,t),
\end{equation}
where $w$ is a constant. Then, a vacuum equation of state has
$w=-1$, while the matter dominated universe has $w=0$. We choose
for simplicity $f(r)=1$ for all values of $r$ and we assume that
the universe is dominated by the cosmological constant with
$p(r,t)=-\rho(r,t)$ and $\sqrt{\Lambda/3} =\sqrt{8\pi G\rho_{\rm
vac}/3}$. For the special case of a spatially flat universe, a
solution to Eqs.(\ref{inhomoF}) and (\ref{inhomoF2}) is given by
\begin{equation}
\label{generaldeSitter}
R(r,t)=R_0\exp\biggl[(\sqrt{\Lambda/3})(r+t)\biggr].
\end{equation}
By a change of radial coordinate
\begin{equation}
{\tilde r}= R_0\exp\biggl[(\sqrt{\Lambda/3})r\biggr],
\end{equation}
we obtain the isotropic and homogeneous de Sitter metric:
\begin{equation}
\label{deSitter}
ds^2=dt^2-\exp\biggl[2(\sqrt{\Lambda/3})t\biggr]d{\tilde r}^2
-\exp\biggl[2(\sqrt{\Lambda/3})\biggr]{\tilde r}^2d\Omega^2.
\end{equation}
Thus, the inhomogeneous expanding bubble can inflate when the
cosmological constant (vacuum energy) dominates the expanding
bubble. The solution (\ref{generaldeSitter}) is a special solution
and a more general inhomogeneous inflating solution should exist
for $f(r)\not=0$, but it does illustrate that an inflationary
solution of Einstein's field equations exists for our more general
inhomogeneous metric.

Assuming that the matter scalar field $\phi$ dominates in the
early universe, and substituting for $\rho_\phi$ and $p_\phi$ in
(\ref{inhomoF}) and (\ref{inhomoF2}), we obtain for $\Lambda=0$:
\begin{equation}
\frac{{\dot R}^2}{R^2}+2\frac{\dot R}{R}\frac{\dot R'}{R'}=8\pi
G\biggl[\frac{1}{2}{\dot\phi}^2+\frac{1}{2}(\vec\nabla)^2\phi+V(\phi)\biggr],
\end{equation}
\begin{equation}
\frac{\ddot{R}}{R}+\frac{1}{3}\frac{\dot{R}^2}{R^2}-\frac{1}{3}\frac{\dot{R}'}{R'}
\frac{\dot{R}}{R} =-\frac{8\pi G}{3}[{\dot\phi}^2-V(\phi)].
\end{equation}
We can obtain an inflationary solution of the form
(\ref{generaldeSitter}) by assuming that $V(\phi) \gg
{\dot\phi}^2$, $(\vec\nabla)^2\phi\sim 0$ and $V(\phi)\sim {\rm
constant}$. However, we are required to make assumptions about the
scalar field $\phi$, namely, that in a local patch of inflation it
only depends on time $t$, so that the wave equation
(\ref{waveequation}) can be solved to give the sub-horizon quantum
fluctuations that seed the growth of structure. This appears to be
an apparently unavoidable fine-tuning of the primordial scalar
matter fields that occurs in generic inflationary models.

\section{Late-Time Matter Dominated Universe}

The late-time matter dominated universe will be pictured as a
large scale perturbation enhancement that is described by an exact
inhomogeneous spherically symmetric solution of Einstein's field
equations. The perturbation enhancement is embedded in a matter
dominated universe that approaches asymptotically an Einstein-de
Sitter universe as $t\rightarrow\infty$. An observer will be
off-center from the origin of coordinates of the spherical
perturbation enhancement.

For the matter dominated Lema\^{i}tre-Tolman-Bondi
(LTB)~\cite{Lemaitre,Tolman,Bondi} model with zero pressure $p=0$
and zero cosmological constant $\Lambda=0$, the Einstein field
equations demand that $R(r,t)$ satisfies
\begin{equation}
\label{Requation} 2R{\dot R}^2+2R(1-f^2)=F(r),
\end{equation}
with $F$ being an arbitrary function of class $C^2$. There exist
three possible solutions depending on whether $f^2 < 1, = 1,
> 1$ and they correspond to elliptic (closed), parabolic (flat),
and hyperbolic (open) cases, respectively.

The proper density of matter can be expressed as
\begin{equation}
\label{density} \rho=\frac{F'}{16\pi GR'R^2}.
\end{equation}
We can solve (\ref{density}) to obtain
\begin{equation}
\Omega-1\equiv\frac{\rho}{\rho_c}-1=\frac{1}{H^2_{\rm
eff}}\biggl(\frac{1-f^2}{R^2}-2\frac{f}{R}\frac{f'}{R'}\biggr),
\end{equation}
where
\begin{equation}
H^2_{\rm eff}=H^2_\perp+2H_\perp H_r,
\end{equation}
is an effective Hubble parameter and we have
\begin{equation}
8\pi G\rho_c=\frac{{\dot R}^2}{R^2}+2\frac{{\dot R}}{R}\frac{{\dot
R}'}{R'}.
\end{equation}
We have the three possibilities for the curvature of spacetime: 1)
$f^2 > 1$ open ($\Omega-1 < 0$), 2) $f^2=1$ flat ($\Omega-1=0$),
$f^2 < 1$ closed ($\Omega-1 > 0$).

Since the WMAP data~\cite{Spergel} shows that the universe is
spatially flat to within a few percent, we shall consider the
globally flat case $f^2(r)=1$. The metric reduces to
\begin{equation}
\label{flatmetric} ds^2=dt^2-R^{'2}(r,t)dr^2-R^2(r,t)d\Omega^2.
\end{equation}
A solution for the matter dominated (w=0) universe is
\begin{equation}
R(r,t)=r[t+\beta(r)]^{2/3},
\end{equation}
and $\beta(r)$ is an arbitrary function of $r$ of class
$C^2$~\cite{Bonnor}. The metric (\ref{flatmetric}) becomes
\begin{equation}
\label{matterdommetric}
ds^2=dt^2-(t+\beta)^{4/3}(Y^2dr^2+r^2d\Omega^2),
\end{equation}
where
\begin{equation}
Y=1+\frac{2r\beta'}{3(t+\beta)},
\end{equation}
and
\begin{equation}
\rho=\frac{1}{6\pi G(t+\beta)^2Y}.
\end{equation}

The arbitrary function $\beta(r)$ can be specified in terms of a
density on some spacelike hypersurface $t=t_0$. The metric and
density are singular on the two hypersurfaces $t+\beta=0$ and
$Y=0$, namely, $t_1=-\beta$ and $t_2=-\beta-2r\beta'/3$,
respectively. The model is only valid for $t > \Sigma(r)\equiv
{\rm Max}[t_1(r),t_2(r)]$, and the hypersurface $t(r)=\Sigma(r)$
defines the big-bang. However, our pressureless model requires
that the surface $t(r)=\Sigma(r)$ describes the surface on which
the universe becomes matter dominated (in the LFRW model this
occurs at $z\sim 10^4$). We observe that even in the spatially
flat LTB model, different parts of the universe can enter the
matter dominated era at different times. Our expanding
inhomogeneous, spherically symmetric bubble is embedded in the
matter dominated metric (\ref{matterdommetric}). For $\beta=0$ and
in the limit $t\rightarrow\infty$ we obtain the Einstein-de Sitter
universe
\begin{equation}
ds^2=dt^2-a^2(t)(dr^2+r^2d\Omega^2),
\end{equation}
where $a(t)=t^{2/3}$. Thus, for $\beta=0$ we obtain the FLRW
model. Moreover, the expanding flat LTB model necessarily evolves
to the homogeneous and isotropic FLRW model for a non-vanishing
density, whatever the initial conditions.

\section{The Inhomogeneous Cosmology Deceleration Parameter}

Let us expand $R(r,t)$ in a Taylor series
\begin{equation}
R(r,t)=R[r,t_0-(t_0-t)] =R(r,t_0)\biggl[1-(t_0-t)\frac{{\dot
R}(r,t_0)}{R(r,t_0)} +\frac{1}{2}(t_0-t)^2\frac{{\ddot
R}(r,t_0)}{R(r,t_0)} -...\biggr]
$$ $$
=R(r,t_0)\biggl[1-(t_0-t)H_{0\perp}-\frac{1}{2}(t_0-t)^2q(r,t_0)H_{0\perp}^2-...\biggr],
\end{equation}
where $t_0$ denotes the present epoch and $H_{0\perp}={\dot
R}(r,t_0)/R(r,t_0)$. Moreover, we have
\begin{equation}
q(r,t_0)=-\frac{{\ddot R}(r,t_0)R(r,t_0)}{{\dot R}^2(r,t_0)}.
\end{equation}
By substituting for ${\ddot R}$ from Eq.(\ref{inhomoF2}), we
obtain
\begin{equation}
\label{inhomodeceleration}
q(r,t_0)=\frac{1}{3}+\frac{4\pi\rho_0(r,t_0)}{3H_{0\perp}^2(r,t_0)}
-\frac{\Lambda}{3H_{0\perp}^2(r,t_0)}-\frac{1}{3}\frac{H_{0r}(r,t_0)}{H_{0\perp}(r,t_0)},
\end{equation}
where $H_{0r}(r,t_0)={\dot R}'(r,t_0)/R'(r,t_0)$.

If we set $H_{0\perp}(r,t_0)=H_{0r}(r,t_0)=H(t_0)$ where
$H(t_0)={\dot a}(t_0)/a(t_0)$, then we obtain the spatially flat
FLRW expression for the deceleration parameter:
\begin{equation}
q=\frac{4\pi\rho_0}{3H_0^2}-\frac{\Lambda}{3H_0^2}=\frac{1}{2}-\Omega_\Lambda.
\end{equation}
We see from (\ref{inhomodeceleration}) that different observers
located in different causally disconnected parts of the sky will
observe different values for the deceleration parameter $q$,
depending upon their location and distance from the center of the
spherically symmetric perturbation enhancement. This can lead to
one form of cosmic variance, because the spatial average of all
the observed values of local physical quantities, including the
deceleration parameter $q$, will have an intrinsic uncertainty.

Let us rewrite (\ref{inhomodeceleration}) in the form
\begin{equation}
\label{decel}
q=\frac{1}{3}+\Omega_{0m\perp}-\Omega_{0\Lambda\perp}-\Omega_{0H\perp},
\end{equation}
where
\begin{equation}
\Omega_{0m\perp}=\frac{4\pi\rho_0(r,t_0)}{3H_{0\perp}^2(r,t_0)},\quad
\Omega_{0\Lambda\perp}=\frac{\Lambda}{3H_{0\perp}^2(r,t_0)},\quad
\Omega_{0H\perp}=\frac{1}{3}\frac{H_{0r}(r,t_0)}{H_{0\perp}(r,t_0)}.
\end{equation}
The variance of $q$ is given by the exact non-perturbative
expression:
\begin{equation}
{\rm var}(q)\equiv\langle q^2-\langle
q\rangle^2\rangle^{1/2}={\overline q},
\end{equation}
where
\begin{equation}
\overline{(...)}=\frac{\int d^3x(...)}{\int d^3x},
\end{equation}
denotes the ensemble average.

We now set $\Lambda=0$ in (\ref{decel}) and obtain
\begin{equation}
\label{decel2} q=\frac{1}{3}+\Omega_{0m\perp}-\Omega_{0H\perp}.
\end{equation}
If we have \begin{equation} \Omega_{0H\perp}
> \frac{1}{3}+\Omega_{0m\perp},
\end{equation}
then the cosmic variance ${\rm var}(q)$ can be negative {\it and
cause the universe to accelerate without a cosmological constant
or dark energy.} If the super-horizon long wave inflationary modes
can generate {\it non-perturbatively} a sufficiently large effect
of $O(1)$, then this can drive ${\rm var}(q)$ to  negative values.

\section{Conditions for Deceleration Parameter}

Hirata and Seljak~\cite{Seljak} have recently raised some critical
questions regarding the Kolbe et al.~\cite{Kolbe} suggestion that
long wave length super-horizon perturbations can cause $q$ to be
negative and not violate the strong energy condition $\rho+3p\geq
0$. They used non-FLRW methods to discuss the deceleration
parameter $q$. They begin by defining a local Hubble value
\begin{equation}
H_1=\frac{1}{3}\nabla_\mu u^\mu,
\end{equation}
and a deceleration parameter
\begin{equation}
q_1=-1-\frac{1}{H_1^2}u^\mu\nabla_\mu H_1.
\end{equation}
Here, $u^\mu\nabla_\mu$ is the Lagrangian proper time derivative
$d/t_{\rm proper}$ associated with a matter particle. The
expansion tensor $\theta_{\mu\nu}$ is related to $H_1$ by
$\theta\equiv {\theta^\mu}_\mu=3H_1$. The matter particles follow
geodesics with $\theta_{\mu\nu}u^\nu=\theta_{\mu\nu}u^\mu=0$. From
the Raychaudhuri equation
\begin{equation}
\frac{d\theta}{dt}=-\frac{\theta^2}{3}-\sigma_{\mu\nu}\sigma^{\mu\nu}
+\omega_{\mu\nu}\omega^{\mu\nu}-R_{\mu\nu}u^\mu u^\nu,
\end{equation}
where $R_{\mu\nu}$ is the Ricci tensor and $\sigma$ and $\omega$
denote the shear and vorticity, respectively, it follows that
\begin{equation}
H_1^2q_1=\frac{1}{3}({\sigma^\mu}_\nu{\sigma^\nu}_\mu-{\omega^\mu}_\nu
{\omega^\nu}_\mu)-\frac{1}{3}R_{\mu\nu}u^\mu u^\nu.
\end{equation}
From Einstein's field equation (\ref{Einstein}) for $\Lambda=0$ we
obtain
\begin{equation}
\label{Hdeceleration}
H_1^2q_1=\frac{1}{3}({\sigma^\mu}_\nu{\sigma^\nu}_\mu-{\omega^\mu}_\nu
{\omega^\nu}_\mu)+\frac{4\pi G}{3}(\rho+3p).
\end{equation}

We have ${\sigma^\mu}_\nu{\sigma^\nu}_\mu > 0$, so that if the
strong energy condition is satisfied, $\rho+3p \geq 0$, and the
vorticity ${\omega^\mu}_\nu=0$, it follows that $q_1 \geq 0$.
Hirata and Seljak now state that in the synchronous comoving gauge
used by Kolbe et al. to calculate the perturbations up to second
order, $\omega=0$, and the deceleration parameter must satisfy
$q_1\geq 0$ when the strong energy condition is satisfied.

In our model the vorticity $\omega\not= 0$ for our exact
inhomogeneous solution that describes the large scale perturbation
enhancement, so it does not necessarily follow that $q_1 \geq 0$.
We do not have to use the synchronous gauge to do calculations of
the non-pertubative inhomogeneity and so, in general, $\omega\not=
0$.

Harati and Seljak show that for a single inflaton inflationary
model, the long wave length fluctuations at super-horizon cannot
give rise to an effect $O(1)$ for ${\rm var}(q)$. However, this
result is model dependent and a more general hybrid inflation
model could lead to infrared modes that can lead to a shifting of
${\rm var}(q)$ to negative values, and explain the acceleration of
the universe without negative pressure dark energy or a
cosmological constant.

If the vorticity $\omega\not= 0$ and the shear $\sigma\sim0$, then
we can consider the possibility that the long wave length
super-horizon inflationary modes can lead to a negative $q$.
However, it can be argued on the basis of standard inflationary
models that due to the large increase of entropy during
re-heating, there will be a large increase of $\rho+3p$ in
(\ref{Hdeceleration}), while the vorticity is not increased
accordingly and, therefore, one may not expect that $q$ can be
driven to negative values by a sufficiently large vorticity value.
But this depends on the re-heating model and may not be a generic
feature of inflation.

Alternatively, one could investigate these issues in a
non-inflationary model, such as the variable speed of light
bimetric gravity theory~\cite{Clayton}, which does predict a scale
invariant spectrum with the spectral index $n_s\sim 0.97$ in
agreement with the WMAP result. The superluminal mechanism that
solves the horizon and flatness problems, namely, that the ratio
$\gamma=c_\gamma/c_g$, where $c_\gamma$ and $c_g$ denote the speed
of light and the speed of gravitational waves, respectively,
becomes unity soon after it reaches a high value in the very early
universe, as the effect of a phase transition with $c_g=c={\rm
constant}$ ($c$ the currently measured speed of light). The
fluctuations that form the seeds of large scale structure are born
super-horizon but they do not lead to the same consequences as
inflation with a large number of e-folds.

\section{Conclusions}

We have shown that a special de Sitter inflationary solution
exists for a spherically symmetric and inhomogeneous expanding
bubble. Further work must be carried out to find more general
inhomogeneous de Sitter-like solutions from our generalized
Friedmann equations.

We have argued that an exact non-perturbative treatment of
inhomogeneities in a cosmology describing a spherically symmetric
expanding universe, embedded in an asymptotic Einstein-de Sitter
universe, can significantly modify local observables such as the
red shift, expansion rate and the luminosity distance when
super-horizon, long wave length modes are taken into account. We
expect that this will hold true even in the case in which only
adiabatic modes are present and spatial gradients and
non-perturbative effects are not ignored.

Care must be taken when arguments are made about the sizes of
effects associated with the super-horizon inflationary modes in
cosmology. According to our results, the various features of the
power spectrum derived from the WMAP data could also be subject to
significant modifications at the non-perturbative level depending
on the model of inflation adopted. The effects of an expansion
parameter $\theta$ and shear and vorticity parameters should be
reinterpreted in our inhomogeneous model together with the effects
of very long wave length super-horizon inflationary modes, before
conclusions are drawn about their relative importance when
compared to an FLRW universe.

The fact that we should consider the effects of super-inflationary
modes in cosmology and that they lead to a cosmic variance for
various local physical quantities, even though these modes cannot
presently be observed, may be inescapable and lead to an intrinsic
uncertainty in our interpretation of the universe.

\vskip 0.2 true in {\bf Acknowledgments} \vskip 0.2 true in This
work was supported by the Natural Sciences and Engineering
Research Council of Canada. I thank Robert Brandenberger, Joel
Brownstein, Martin Green and Moshe Rozali for helpful
discussions.\vskip 0.5 true in


\begin{thebibliography}{100}
\bibitem{Moffat} J. W. Moffat, astro-ph/0502110.

\bibitem{Guth} A. Guth, Phys. Rev. {\bf D23}, 347 (1981); A. Linde,
Phys. Lett. {\bf B108}, 389 (1982); A. Albrecht and P. J.
Steinhardt, Phys. Rev. Lett. {\bf 48}, 1220 (1982); D. H. Lyth and
A. Riotto, Phys. Rep. {\bf 314}, 1 (1999); W. Kinney,
astro-ph/0301448.

\bibitem{Trodden} M. Trodden and T. Vachaspati, Phys. Rev. {\bf D
61}, 023502 (2000), gr-qc/9811037; Mod. Phys. Lett. {\bf A14},
1661 (1999), gr-qc/9905091; R. H. Brandenberger and J. H. Kung,
Phys. Rev. {\bf D42}, 1008 (1990).

\bibitem{Perlmutter} S. Perlmutter et al. Ap. J. {\bf 517},
565 (1999); A. G. Riess, et al. Astron. J. {\bf 116}, 1009 (1998);
P. M. Garnavich, et al. Ap. J. {\bf 509}, 74 (1998).

\bibitem{Riess} A. G. Riess, et al., Ap. J. {\bf 607}, 665 (2004),
astro-ph/0402512.

\bibitem{Spergel} C. L. Bennett et al., Ap. J. Suppl. {\bf 148}, 1
(2003), astro-ph/0302207; D. N. Spergel et al., Ap. J. Suppl. {\bf
148}. 175 (2003), astro-ph/0302209.

\bibitem{Turner} See e.g. S. M. Carrroll, V. Duvvuri, M. Trodden
and M. Turner, Phys. Rev. {\bf D70}, 043528 (2004).

\bibitem{Peebles} P. J. E. Peebles and B. Ratra, Rev. Mod. Phys.
{\bf 75}, 559 (2003); N. Straumann, astro-ph/0203330.

\bibitem{Kolbe} E. W. Kolbe, S. Matarrese, A. Notari and A.
Riotto, Phys. Rev. {\bf D71}, 023524 (2004); {\it ibid},
astro-ph/0410541; {\it ibid}, hep-th/0503117.

\bibitem{Barausse} E. Barausse, S. Matarrese and A. Riotto,
astro-ph/05001152.

\bibitem{Brandenberger} V. F. Mukhanov, L. R. W. Abramo and R. H.
Brandenberger, Phys. Rev. Lett. {\bf 78}, 1624 (1997); W. Unruh,
astro-ph/9802323; N. Ashfordi and R. H. Brandenberger, Phys. Rev.
{\bf D63}, 123505 (2001); G. Geshnizjani and R. H. Brandenberger,
hep-th/0310265; R. H. Brandenberger and C. S. Lam, hep-th/0407048;
N. C. Tsamis and R. P. Woodard, Nucl. Phys. {\bf B474}, 235
(1996), hep-ph/9602315; Y. Nambu, gr-qc/0503111.

\bibitem{Chung} G. Geshnizjani and D. J. H. Chung,
astro-ph/0503553.

\bibitem{Flanagan} E. E. Flanagan, hep-th/0503202.

\bibitem{Seljak} C. M Hirata and U. Seljak, astro-ph/0503582.

\bibitem{Wiltshire} D. L. Wiltshire, gr-qc/0503099.

\bibitem{Lemaitre} G. Lema\^{i}tre, Mon. Not. Roy. Astron. Soc. {\bf 91},
490 (1931); Ann. Soc. Sci. Bruxelles, {\bf A53}, 51 (1933).

\bibitem{Tolman} R. C. Tolman, Proc. Nat. Acad. Sci. {\bf 20},
169 (1934).

\bibitem{Bondi} H. Bondi, Mon. Not. Roy. Astron. Soc. {\bf 107},
410 (1947).

\bibitem{Bonnor} W. B. Bonnor, Mon. Not. Roy. Astron. Soc. {\bf 167},
55 (1974).

\bibitem{Moffat2} J. W. Moffat and D. C. Tatarski, Ap. J.
{\bf 453}, 17 (1990).

\bibitem{Moffat3} J. W. Moffat and D. C. Tatarski, Phys. Rev.
{\bf D45}, 3512 (1992).

\bibitem{Krasinski} For a review of inhomogeneous cosmology and
references, see: A Krasi\'nski, {\it Inhomogeneous Cosmological
Models}, Cambridge University Press, 1997.

\bibitem{Clayton} M. A. Clayton and J. W. Moffat, JCAP 0307 (2003)
004, gr-qc/0304058; J. W. Moffat, 32nd Coral Gables Conference,
Fort Lauderdale, Florida, December 17-21, 2003. Published in the
Proceedings of the Conference, Eds. T. Curtright, S. Mintz and A.
Perlmutter, World Scientific, 2004, gr-qc/0404066.


\end{thebibliography}
\end{document}